\documentstyle[graphicx, 12pt]{article}
\begin{document}

\small
\hoffset=-1truecm
\voffset=-2truecm
\title{\bf The Casimir effect at finite temperature in a six-dimensional
vortex scenario}
\author{Hongbo Cheng\footnote {E-mail address: hbcheng@ecust.edu.cn}\\
Department of Physics, East China University of Science and
Technology,\\ Shanghai 200237, China}

\date{}
\maketitle

\begin{abstract}
The Casimir effect for parallel plates satisfying the Dirichlet
boundary condition in the context of effective QED coming from a
six-dimensional Nielsen-Olesen vortex solution of the Abelian
Higgs model with fermions coupled to gravity is studied at finite
temperature. We find that the sign of the Casimir energy remains
negative under the thermal influence. It is also shown that the
Casimir force between plates will be weaker in the
higher-temperature surroundings while keeps attractive. This
Casimir effect involving the thermal influence is still
inconsistent with the known experiments. We find that the thermal
correction can not compensate or even reduce the modification from
this kind of vortex model to make the Casimir force to be in less
conflict with the measurements.
\end{abstract}
\vspace{7cm} \hspace{1cm} PACS number(s): 03.70.+k, 03.65.Ge

\newpage

\noindent \textbf{I.\hspace{0.4cm}Introduction}

In order to unify the interactions in nature and solve the
long-standing problems such as the hierachy problem, the
cosmological constant problem, etc. from particle physics and
cosmology, various models of higher-dimensional spacetime were put
forward and attract a lot of attention [1-13]. Within the frame of
field theory for the approaches with more than four dimensions,
the mechanisms for localization of the fields like scalar field,
fermionic field and gauge field on the models must be needed, then
these models can describe the matter fields and interactions
[14-19]. It should be pointed out that an effective quantum
electrodynamics in four-dimensional spacetime generated from a
Higgs model with fermions coupled to gravity in the world with six
dimensions lead the fermionic and gauge functions spread on the
transverse direction in a small region around the core of a
Nielsen-Olesen vortex [16]. In this issue the localization of
gauge field was achieved and the Nielsen-Olesen vortex comes from
the fluctuations of graviton and gauge field. It is significant
that this construction admits the gravity and gauge field besides
the scalar and fermionic fields.

The Casimir effect is essentially a direct consequence of quantum
field theory subject to a change in the spacetime of vacuum
oscillations under the inserted background field [20-26]. It is
important to discuss the sign of the Casimir energy and the nature
of the Casimir force for many subjects because the Casimir effect
depends on various factors. The precision of the measurements has
been greatly improved experimentally [27]. We can compare the
theoretical results with the phenomena to explore the properties
of various worlds. The Casimir effect has opened a new window to
explore the topics. The Casimir effect for parallel plates in the
spacetime with extra compactified dimensions was discussed
[28-38]. The distinct influence from the fractal additional
compactified dimension can be exhibited in the Casimir effect for
parallel plates [39, 40]. The Casimir effect for the same system
in the braneworld was also studied [41-50].

It is significant to research on the Casimir effect for parallel
plates within the frame of the effective QED. The new QED
developing from a six-dimensional Abelian Higgs model coupled to
gravity in a Nielsen-Olesen vortex background with fermions is a
mixture of the original six-dimensional metric and the vector
potential [16]. The coupling constant and the size of the core of
the six-dimensional vortex as the own features of this kind of
effective gauged field theory modify the standard Casimir effect
between parallel plates obviously [51]. The manifest deviations
are not consistent with the known experiments, which making the
phenomenological viability of the model less [51]. The works on
the modified Casimir force can be used to explore the
four-dimensional effective QED in a new direction [51].

The quantum field at finite temperature shares many effects. In
many cases the thermal influence on the Casimir effect cannot be
neglected, and its influence certainly modifies the effect. The
Casimir effect for parallel plates under a nonzero temperature
environment in the presence of additional compactified dimensions
was considered and the magnitude of Casimir force as well as the
sign of Casimir energy change with the temperature [53-55]. The
Casimir energy for a rectangular cavity including thermal
corrections was considered and the temperature controls the energy
sign [56]. The Casimir effect for a scalar field within two
parallel plates under thermal influence in the bulk region of
Randall-Sundrum models was also evaluated [57, 58]. In addition,
the thermal modification to the Casimir effect for parallel plates
involving massless Majorana fermions was analyzed [59].

It is also fundamental to probe the Casimir effect for
parallel-plate system at finite temperature in the context of the
effective QED. To our knowledge little contribution was made to
this topic. Now we plan to study the Casimir effect for parallel
plates in the background of a six-dimensional vortex scenario to
generalize some of the conclusions of Ref. [51]. We wonder how the
thermal influence modifies the Casimir effect in the vortex
scenario and the difference between the effect limited by the
higher-dimensional vortex and the standard one in particular. At
first we derive the frequency of massless scalar field subject to
the Dirichlet boundary conditions on the plates with thermal
corrections in a Nielsen-Olesen vortex background by means of
finite-temperature field theory. We regularize the frequency to
obtain the Casimir energy density with the help of the zeta
function technique and the Casimir force between the parallel
plates further. We will compare our results with the standard
Casimir effect to determine how the temperature change their
difference. Our discussions and conclusions are listed at the end
of this paper.

\vspace{0.8cm} \noindent \textbf{II.\hspace{0.4cm}The Casimir
effect at finite temperature in a six-dimensional vortex scenario}

We make use of the imaginary time formalism to describe the Higgs
field and gauge field in thermal equilibrium to introduce a
partition function for a parallel-plate device [52],

\begin{equation}
Z=N'\int_{periodic}DA_{\mu}D\Phi\exp[\int_{0}^{\beta}d\tau\int
d^{5}x\sqrt{-g}\mathcal{L}(\Phi, A^{\mu}, \partial_{E}\Phi,
\partial_{E}A^{\mu})]
\end{equation}

\noindent where $\mathcal{L}$ is the Lagrangian density for Higgs
field and gauge field within the parallel-plate structure under
consideration. $N'$ is a constant and "periodic" means [52],

\begin{equation}
\Phi(0, x^{i})=\Phi(\tau=\beta, x^{i})
\end{equation}

\begin{equation}
A^{\mu}(0, x^{i})=A^{\mu}(\tau=\beta, x^{i})
\end{equation}

\noindent where $i=1, 2, 3, 4, 5$. $\beta=\frac{1}{T}$ is the
inverse of the temperature and $\tau=it$ The scalar field
satisfies the Dirichlet boundary condition on the plates to lead
the wave vector in the direction restricted by the plates to be
$k_{N}=\frac{\pi N}{L}$, where $L$ is the plate separation.
According to the solution to the field equations of effective QED
[16, 51] and the boundary conditions on the fields, the
generalized zeta function can be written as [52],

\begin{eqnarray}
\zeta(s,
-\partial_{E})=Tr(-\partial_{E})^{-s}\hspace{7.5cm}\nonumber\\
=\frac{1}{2}ap\int\frac{d^{2}k_{\perp}dk_{r}}{(2\pi)^{3}}
\sum_{n,N=1}^{\infty}\sum_{q=-\infty}^{\infty}[k_{\perp}^{2}
+\frac{2}{e}k_{r}^{2}+\frac{\pi^{2}N^{2}}{L^{2}}
+\frac{2}{el^{2}}\frac{1}{n^{2}}+(\frac{2q\pi}{\beta})^{2}]^{-s}
\end{eqnarray}

\noindent where

\begin{equation}
\partial_{E}=\frac{\partial^{2}}{\partial\tau^{2}}
-g^{ij}\partial_{i}\partial_{j}
\end{equation}

\noindent leading the dispersion relation shown in Eq. (4)
involving not only the terms from 4-dimensional spacetime but also
the contributions from the additional compactified space and
$g_{\mu\nu}$ with $\mu,\nu=0,1,2,3,4,5$ is the metric of
six-dimensional vortex scenario [16, 18]. Here $k_{\perp}$ denotes
the two transverse components of the momentum. In the
two-dimensional additional space the continuum variable $k_{r}$
comes from the radial part and the other component coming from the
vortex number appear as the term $\frac{2}{el^{2}}\frac{1}{n^{2}}$
with $l=\frac{\kappa}{e}$ the ratio of the six-dimensional
gravitational constant and the parameter of the Abelian Higgs
model [51]. Here the parameter $a$ standing for the size of the
vortex'es core is due to the integration in $k_{r}$ [51]. The
factor $p$ represents the possible polarization of the photon
[51]. We make use of
$\varepsilon=-\frac{\partial}{\partial\beta}(\frac{\partial\zeta(s;-\partial_{E})}
{\partial s}|_{s=0})$ [52] to obtain the total energy density of
the system with thermal corrections as follow,

\begin{equation}
\varepsilon=\frac{1}{4\pi^{2}}ap\sqrt{\frac{e}{2}}\Gamma(\frac{3}{2})
\Gamma(-\frac{3}{2})\frac{\partial}{\partial\beta}M_{3}(-\frac{3}{2};
\frac{2}{el^{2}},\frac{\pi^{2}}{L^{2}},\frac{4\pi^{2}}{\beta^{2}};
2,2,-2)
\end{equation}

\noindent where the multiple zeta functions with arbitrary
exponents is [23, 24],

\begin{equation}
M_{N}(s;a_{1},a_{2},\cdot\cdot\cdot,a_{N};\alpha_{1},\alpha_{2},
\cdot\cdot\cdot,\alpha_{N})=\sum_{n_{1},n_{2},\cdot\cdot\cdot,n_{N}=1
}^{\infty}(a_{1}n_{1}^{\alpha_{1}}+a_{2}n_{2}^{\alpha_{2}}
+\cdot\cdot\cdot+a_{N}n_{N}^{\alpha_{N}})^{-s}
\end{equation}

\noindent With the help of the zeta function technique [23, 24],
we regularize the total energy density to obtain the Casimir
energy per unit area for parallel plates at finite temperature,

\begin{eqnarray}
\varepsilon_{C}=\frac{\pi^{-\frac{5}{2}}}{32}ap\sqrt{\frac{e}{2}}
(\frac{2}{el^{2}})^{2}\Gamma(\frac{3}{2})\Gamma(-2)\zeta(4)
-\frac{\pi^{-3}}{32}ap\sqrt{\frac{e}{2}}(\frac{2}{el^{2}})^{\frac{5}{2}}
L\Gamma(\frac{3}{2})\Gamma(-\frac{5}{2})\zeta(5)\nonumber\\
-\frac{\pi^{-3}}{8}ap\sqrt{\frac{e}{2}}(\frac{2}{el^{2}})^{\frac{5}{4}}
L^{-\frac{3}{2}}\Gamma(\frac{3}{2})\sum_{n_{1},n_{2}=1}^{\infty}
(\frac{1}{n_{1}n_{2}})^{\frac{5}{2}}K_{\frac{5}{2}}(2\sqrt{\frac{2}{e}}
\frac{L}{l}\frac{n_{1}}{n_{2}})\hspace{1cm}\nonumber\\
+\pi^{-\frac{5}{2}}\Gamma(\frac{3}{2})ap\sqrt{\frac{e}{2}}\frac{1}{\beta^{2}}
\sum_{n_{1},n_{2},n_{3}=1}^{\infty}n_{1}^{-2}(\frac{\pi^{2}}{L^{2}}n_{2}^{2}
+\frac{2}{el^{2}}\frac{1}{n_{3}^{2}})\hspace{2.5cm}\nonumber\\
\times K_{2}(\beta
n_{1}\sqrt{\frac{\pi^{2}}{L^{2}}n_{2}^{2}+\frac{2}{el^{2}}\frac{1}{n_{3}^{2}}})
\hspace{3cm}\nonumber\\
+\frac{\pi^{-\frac{5}{2}}}{2}\Gamma(\frac{3}{2})ap\sqrt{\frac{e}{2}}
\frac{1}{\beta}\sum_{n_{1},n_{2},n_{3}=1}^{\infty}n_{1}^{-1}
(\frac{\pi^{2}}{L^{2}}n_{2}^{2}+\frac{2}{el^{2}}\frac{1}{n_{3}^{2}})^{\frac{3}{2}}
\hspace{2cm}\nonumber\\
\times[K_{1}(\beta
n_{1}\sqrt{\frac{\pi^{2}}{L^{2}}n_{2}^{2}+\frac{2}{el^{2}}\frac{1}{n_{3}^{2}}})
+K_{3}(\beta
n_{1}\sqrt{\frac{\pi^{2}}{L^{2}}n_{2}^{2}+\frac{2}{el^{2}}\frac{1}{n_{3}^{2}}})]
\nonumber\\
<0\hspace{12cm}
\end{eqnarray}

\noindent where $K_{\nu}(z)$ is the modified Bessel function of
the second kind. In the expression of Casimir energy like Eq. (8)
the terms with series converge very quickly and only the first
several summands need to be taken into account for numerical
calculation to further discussion. If the temperature approaches
zero, the Casimir energy will recover to be the results of Ref.
[51], so will the Casimir force. The Casimir energy remains
negative no matter how strong the thermal influence is.

In the six-dimensional vortex background the Casimir force on the
plates is also obtained by the derivative of the Casimir energy
with respect to the plate distance. The Casimir force per unit
area on the plates due to the Dirichlet boundary condition
becomes,

\begin{equation}
f'_{C}=\frac{1}{32\pi^{3}}ap\sqrt{\frac{e}{2}}(\frac{2}{el^{2}})^{\frac{5}{2}}
\Gamma(\frac{3}{2})\Gamma(-\frac{5}{2})\zeta(5)+f_{C}
\end{equation}

It should be pointed out that the first term in the expression
above also has nothing to do with the plates gap, which means that
there are two forces with the same magnitude and the opposite
direction acting on the plates respectively, so the two forces
will be compensated each other on every plate. We can remove this
term to write the net Casimir force as follow,

\begin{eqnarray}
f_{C}=-\frac{3}{16\pi^{3}}\Gamma(\frac{3}{2})ap\sqrt{\frac{e}{2}}
(\frac{2}{el^{2}})^{\frac{5}{4}}\frac{1}{L^{\frac{5}{2}}}
\sum_{n_{1},n_{2}=1}^{\infty}(\frac{1}{n_{1}n_{2}})^{\frac{5}{2}}
K_{\frac{5}{2}}(2\sqrt{\frac{2}{e}}\frac{L}{l}\frac{n_{1}}{n_{2}})
\hspace{1cm}\nonumber\\
-\frac{1}{8\pi^{3}}\Gamma(\frac{3}{2})ap\sqrt{\frac{e}{2}}(\frac{2}{el^{2}})^{\frac{7}{2}}
\frac{1}{L^{\frac{3}{2}}}\sum_{n_{1},n_{2}=1}^{\infty}
n_{1}^{-\frac{3}{2}}n_{2}^{-\frac{7}{2}}\hspace{2.5cm}\nonumber\\
\times[K_{\frac{3}{2}}(2\sqrt{\frac{2}{e}}\frac{L}{l}\frac{n_{1}}{n_{2}})
+K_{\frac{7}{2}}(2\sqrt{\frac{2}{e}}\frac{L}{l}\frac{n_{1}}{n_{2}})]
\hspace{1cm}\nonumber\\
+\frac{2}{\pi^{\frac{1}{2}}}\Gamma(\frac{3}{2})ap\sqrt{\frac{e}{2}}
\frac{1}{\beta^{2}}\frac{1}{L^{3}}\sum_{n_{1},n_{2},n_{3}=1}^{\infty}
(\frac{n_{2}}{n_{1}})^{2}K_{2}(\beta
n_{1}\sqrt{\frac{\pi^{2}}{L^{2}}n_{2}^{2}+\frac{2}{el^{2}}\frac{1}{n_{3}^{2}}})\nonumber\\
+\frac{1}{\pi^{\frac{1}{2}}}\Gamma(\frac{3}{2})ap\sqrt{\frac{e}{2}}
\frac{1}{\beta}\frac{1}{L^{3}}\sum_{n_{1},n_{2},n_{3}=1}^{\infty}
\frac{n_{2}^{2}}{n_{1}}(\frac{\pi^{2}}{L^{2}}n_{2}^{2}
+\frac{2}{el^{2}}\frac{1}{n_{3}^{2}})^{\frac{1}{2}}\hspace{2cm}\nonumber\\
\times[K_{1}(\beta n_{1}\sqrt{\frac{\pi^{2}}{L^{2}}n_{2}^{2}
+\frac{2}{el^{2}}\frac{1}{n_{3}^{2}}})+K_{3}(\beta
n_{1}\sqrt{\frac{\pi^{2}}{L^{2}}n_{2}^{2}
+\frac{2}{el^{2}}\frac{1}{n_{3}^{2}}})]\nonumber\\
-\frac{1}{4\pi^{\frac{1}{2}}}\Gamma(\frac{3}{2})ap\sqrt{\frac{e}{2}}
\frac{1}{L^{3}}\sum_{n_{1},n_{2},n_{3}=1}^{\infty}\frac{n_{2}^{2}}{n_{1}}
(\frac{\pi^{2}}{L^{2}}n_{2}^{2}
+\frac{2}{el^{2}}\frac{1}{n_{3}^{2}})\hspace{2cm}\nonumber\\
\times[K_{0}(\beta n_{1}\sqrt{\frac{\pi^{2}}{L^{2}}n_{2}^{2}
+\frac{2}{el^{2}}\frac{1}{n_{3}^{2}}})+2K_{2}(\beta
n_{1}\sqrt{\frac{\pi^{2}}{L^{2}}n_{2}^{2}
+\frac{2}{el^{2}}\frac{1}{n_{3}^{2}}})\nonumber\\
+K_{4}(\beta n_{1}\sqrt{\frac{\pi^{2}}{L^{2}}n_{2}^{2}
+\frac{2}{el^{2}}\frac{1}{n_{3}^{2}}})]\hspace{4cm}
\end{eqnarray}

\noindent The nature of the net Casimir force appearing as the
Casimir effect is attractive. Similarly, if the temperature
vanishes, the Casimir pressure shown in Eq. (10) will be
consistent with that of Ref. [51]. When the two plates are moved
farther and farther away from each other, the Casimir effect will
disappear gradually,

\begin{equation}
\lim_{L\longrightarrow\infty}f_{C}=0
\end{equation}

\noindent It is also evident that the Casimir force is weaker and
weaker with stronger thermal influence like,

\begin{equation}
\lim_{T\longrightarrow\infty}f_{C}=0
\end{equation}

\noindent The asymptotic behaviours of the Casimir force in the
cases of two parallel plates locating remotely each other or
higher temperature respectively are favoured experimentally [27].
The Casimir energy and the Casimir force are modified by the
parameters of six-dimensional vortex as well as the temperature,
but the influence from the topological defect works as a
multiplicative factor on the Casimir effect involving the thermal
correction rather than only term added in the expressions, which
is similar to the results in Ref. [51]. The terms associated with
the temperature are also multiplied by the model variables and can
not diminish the vortex correction independently. In the
environment of six-dimensional vortex, the shapes of the net
Casimir force between two parallel plates subject to the various
temperature are similar although these thermal modifications are
manifest. The factors showing the gauge coupling and the size of
the vortex with six dimensions are dominant to the whole Casimir
effect, but the evident deviations from these factors are
inconsistent with the observational results [27, 51] and the
thermal influence can not cancel and even reduce the
experimentally exclusive deviations.

\vspace{0.8cm} \noindent \textbf{III.\hspace{0.4cm}Discussion}

In this paper the Casimir effect with thermal correction for
parallel plates is investigated in the frame of six-dimensional
Nielsen-Olesen vortex with fermions coupling to gravity. At first
we obtain the Casimir energy for parallel-plate system satisfying
the Dirichlet boundary condition under the vortex-controlled
environment with more than four dimensions at finite temperature
and find that the energy remains negative no matter how the
temperature changes. It is also found that the magnitude of the
Casimir force between two parallel plates with the same boundary
conditions will be weaker as the surroundings are hotter while the
Casimir force keeps attractive. As the temperature is extremely
high, the Casimir force on the plates will vanish. The Casimir
force between the plates will approach zero when the two plates
are moved apart from each other much farther. It should be pointed
out that the thermal influence modifies the Casimir effect such as
the magnitude of the Casimir force and even makes the force nearly
disappear at extremely high temperature, but the temperature can
not greatly weaken or cancel the effect from the multiplicative
factor of the effective QED describing the higher-dimensional
Nielsen-Olesen vortex with fermions coupling to gravity. Although
the Casimir effect we consider here contains the thermal
corrections, the parameters of the effective QED control the shape
of the Casimir effect, so the thermal influence can not improve
the phenomenological viability of the model.

\vspace{1cm}
\noindent \textbf{Acknowledge}

This work is supported by NSFC No. 10875043.

\end{document}